\documentclass[11pt,osajnl2,showpacs,superscriptaddress]{revtex4}
\usepackage{hyperref}
\usepackage{epsfig,rotating}
\usepackage{natbib}
\renewcommand{\baselinestretch}{1.5}

\DeclareRobustCommand{\baselinestretch{1.0}}

\font\boldsym=cmmib10
\def    \aeff   {a_{\rm eff}}
\def    \ATUC	{A_{\rm TUC}}
\def    \balpha {{\hbox{\boldsym\char'013}}}    

\def    \bA     {{\bf A}}
\def    \bAtilde {{\bf\tilde{A}}}
\def    \bB     {{\bf B}}
\def    \bE     {{\bf E}}
\def    \bF     {{\bf F}}

\def    \bP     {{\bf P}}

\def    \bR     {{\bf R}}
\def    \bS     {{\bf S}}
\def    \bT     {{\bf T}}

\def    \bk     {{\bf k}}
\def    \bLu     {{\bf L}_u}
\def    \bLv     {{\bf L}_v}

\def    \br     {{\bf r}}
\def    \bu     {{\bf u}}
\def    \bv     {{\bf v}}

\def    \Csca   {C_{\rm sca}}

\def    \ehat  {\hat{\bf e}}
\def    \FTUC   {\bF_{\rm TUC}}
\def    \inc    {{\rm inc}}
\def    \khat  {\hat{\bf k}}

\def    \ndim   {{\nu}}
\def    \Pbar   {{\bar{P}}}
\def    \phihat {\hat{\bf\phi}}
\def    \polang {{\zeta}}
\def    \RF     {R_{F}}           
\def    \thetahat {\hat{\bf\theta}}

\def    \xhat  {\hat{\bf x}}
\def    \yhat  {\hat{\bf y}}

\newcommand{\ltsim}{\lesssim}
\newcommand{\gtsim}{\gtrsim}

\def    \beq    {\begin{equation}}
\def    \eeq    {\end{equation}}
\def    \beqa    {\begin{eqnarray}}
\def    \eeqa    {\end{eqnarray}}

\countdef\decade=200
\advance\decade by \year
\countdef\hours=201
\advance\hours by \time
\divide\hours by 60
\countdef\mins=202
\advance\mins by \hours
\multiply\mins by 60
\multiply\hours by 100
\countdef\miltime=203
\def   \josaa     {J.~Opt.~Soc.~Am.~A} 
\def   \optlett   {Optics~Lett.}

\begin{document}
\title{The discrete dipole approximation for periodic targets:
		I. theory and tests%
       }
\vspace*{-4em}
\centerline{{\it J. Optical Society of America, A}, submitted}
\vspace*{3em}
\author{B. T. Draine}
\email{draine@astro.princeton.edu}
\affiliation{Princeton University Observatory}
\author{Piotr J. Flatau}
\affiliation{Scripps Institution of Oceanography, University of California, San Diego}

\date{\today}

\begin{abstract}
The discrete-dipole approximation (DDA) is a powerful method for
calculating absorption and scattering by targets that have sizes
smaller than or comparable to the wavelength of the incident radiation.  
The DDA can be extended to targets that are singly- or doubly-periodic.
We generalize the scattering amplitude matrix and the $4\times4$ Mueller matrix
to describe scattering by singly- and doubly-periodic targets, and
show how these matrices can be calculated using the DDA.
The accuracy of DDA calculations using the open-source code DDSCAT
is demonstrated by comparison to exact results for infinite
cylinders and infinite slabs.
A method for using the DDA solution to obtain fields 
within and near the target is presented,
with results shown for infinite slabs.
\end{abstract}


\maketitle
\section{Introduction}

Electromagnetic scattering 
is used to study isolated particles, but increasingly to characterize
extended targets ranging from
nanostructure arrays in laboratories, to
planetary and asteroidal regoliths.
To model the absorption and scattering,
Maxwell's equations must be solved for the target geometry.

For scattering by isolated particles with complex geometry, 
a number of different theoretical approaches have been used,
including the discrete dipole 
approximation (DDA) 
\cite{Purcell+Pennypacker_1973,
      Draine_1988,
      Draine+Flatau_1994,
      Draine_2000a},
also known as the coupled dipole approximation
or coupled dipole method.
The DDA can treat inhomogeneous targets and anisotropic materials, and
has been extended to treat targets near substrates
\cite{Schmehl+Nebeker+Hirleman_1997,Paulus+Martin_2001}.
Other techniques have also been employed, including
the finite difference time domain (FDTD) method
\cite{Yang+Liou_2000,Taflove+Hagness_2005}.


For illumination by monochromatic plane waves, 
the DDA can be extended to targets
that are spatially periodic and (formally) infinite in extent.  
This could apply, for example, to
a periodic array of nanostructures in a laboratory setting, 
or it might be
used to approximate a regolith by a periodic array of
``target unit cells'' with complex structure within each unit cell.

Generalization of the DDA 
(= coupled dipole method) 
to periodic structures was first
presented by
Markel \cite{Markel_1993} for a 1-dimensional chain of dipoles,
and more generally by 
Chaumet et al.\ \cite{Chaumet+Rahmani+Bryant_2003},
who calculated the electric field near
a 2-dimensional array of parallelepipeds illuminated by a plane wave.
Chaumet and Sentenac \cite{Chaumet+Sentenac_2005} 
further extended the DDA 
to treat periodic structures with a finite number of defects. 

From a computational standpoint, solving
Maxwell's equations for periodic structures is 
only slightly more difficult than calculating
the scattering properties of a single ``unit cell'' from the structure.
Since it is now feasible to treat targets with $N\gtsim 10^6$ dipoles
(target volume $\gtsim 200 \lambda^3$, where $\lambda$ is the
wavelength), it becomes possible
to treat extended objects with complex substructure.

The objective of the present paper is to present the theory 
of the DDA applied to 
scattering and absorption by structures that are periodic in one or two
spatial dimensions.
We also generalize the standard formalism for describing the far-field
scattering properties of finite targets (the $2\times2$ scattering
amplitude matrix, and $4\times4$ Mueller matrix) to describe scattering
by periodic targets.
We show how to calculate
the Mueller matrix to describe scattering of
arbitrarily-polarized radiation.
The theoretical treatment developed here has been implemented in 
the open-source code DDSCAT 7 (see Appendix A).

The theory of the DDA for periodic targets is reviewed in 
\S\ref{sec:dda for periodic targets}, and the formalism for describing the
far-field scattering properties of periodic targets is presented in 
\S\ref{sec:in the radiation zone}-\ref{sec:mueller matrix}.
Transmission and reflection coefficients for targets that are periodic
in two dimensions are obtained in \S\ref{sec:T and R}.

The applicability and accuracy of the DDA method
are discussed in \S\S\ref{sec:cylinder},\ref{sec:slab}, where we
show scattering properties calculated
using DDSCAT 7
for two
geometries for which exact solutions are available for comparison:
(1) an infinite cylinder 
and (2)
an infinite slab of finite thickness.
The numerical comparisons demonstrate that, for given $\lambda$, 
the DDA converges to the
exact solution as the interdipole spacing $d\rightarrow0$.

\section{\label{sec:dda for periodic targets}
         DDA for Periodic Targets}

The discrete-dipole approximation (DDA) is a general technique for
calculating scattering and absorption of electromagnetic radiation
by targets with arbitrary geometry.
The basic theory of the DDA has been presented elsewhere
\cite{Draine+Flatau_1994}.
Conceptually, the DDA consists of approximating the target of interest
by an array of polarizable points, with specified polarizabilities.
Once the polarizabilities are specified, Maxwell's equations can
be solved accurately for the dipole array.  When applied to finite
targets, the DDA is limited by the number of dipoles $N$ for which 
computations are feasible -- the limitations may arise from large
memory requirements, or the large amount of computing that may be
required to find a solution when $N$ is large.  In practice, modern
desktop computers are capable of solving the DDA equations, as implemented
in DDSCAT 
\cite{Draine+Flatau_1994,Draine+Flatau_2004}, 
for $N$ as large as $\sim10^6$.

\begin{figure}[t]
\includegraphics[angle=0,width=3.25in]{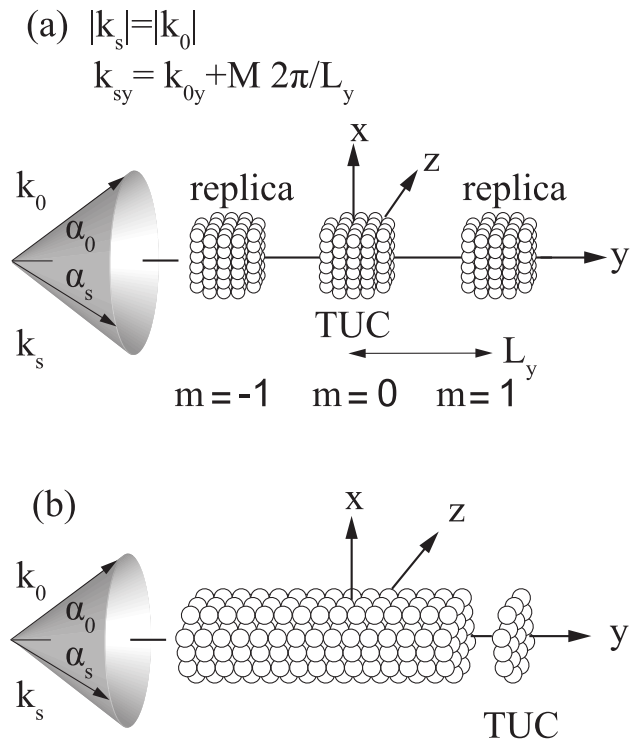} 
\caption{\label{fig:1d} 
         (a) Target consisting of a 1-d array of TUCs
         and (b) showing how
         an infinite cylinder can be constructed from disk-like TUCs (lower).
         The $M=0$ scattering cone, with $\alpha_s=\alpha_0$, is illustrated.}
\end{figure}

\begin{figure}[t]
\includegraphics[angle=0,width=3.25in]{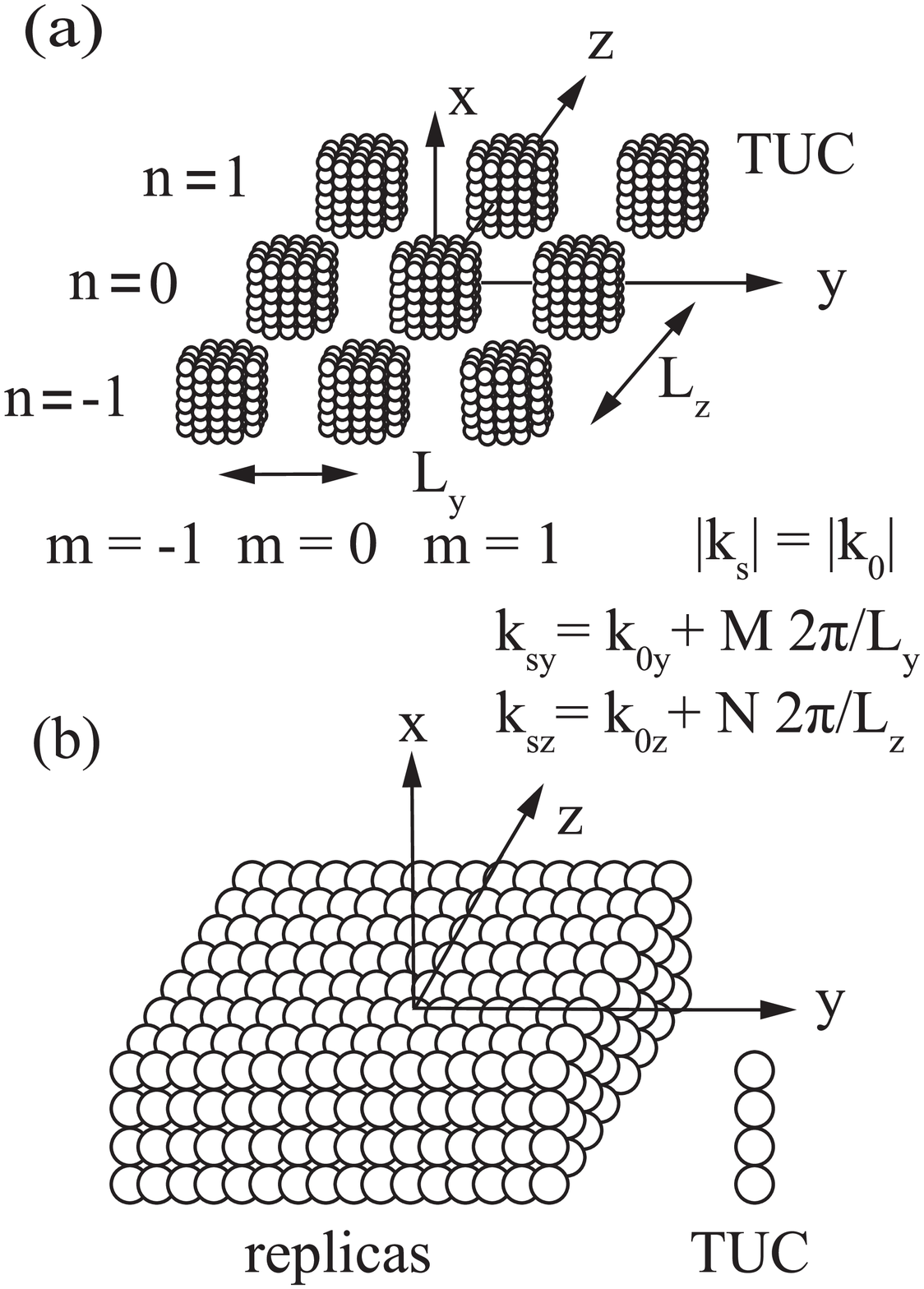}  
\caption{\label{fig:2d} 
         (a) Target consisting of a 2-d array of TUCs and
         (b) showing how an infinite slab is created from
             TUCs consisting of a single ``line'' of dipoles.}
\end{figure}

Developed originally to study scattering from isolated, finite
structures such as dust grains 
\cite{Purcell+Pennypacker_1973}, 
the DDA can be extended to treat singly- or
doubly-periodic structures.
Consider a collection of $N$ polarizable points, defining 
a ``target unit cell'' (TUC).
Now consider a target consisting of a 1-dimensional or 2-dimensional
periodic array of identical TUCs, as illustrated in
Figs.\ \ref{fig:1d} and \ref{fig:2d}; we will refer to these as 1-d
or 2-d targets, although the constituent TUC
may have an arbitrary 3-dimensional shape.
For a monochromatic
incident plane wave
\beq \label{eq:Einc}
\bE_{\rm inc}(\br,t) = \bE_0 \exp\left(i\bk_0\cdot \br -i\omega t\right) ~~~,
\eeq
the polarizations of the dipoles in the target will oscillate coherently.
Each dipole will be affected by the incident wave plus the electric field
generated by {\it all} of the other point dipoles.

Let index $j=1,...,N$ run over the dipoles in a single TUC,
and 
let indices $m$, $n$ run over replicas of the TUC.
The $(m,n)$ replica 
of dipole $j$ is located at
\beq \label{eq:P periodicity}
\br_{jmn} = \br_{j00} + 
m\bLu + n \bLv
~,~~
\eeq
where $\bLu$ and $\bLv$ are the lattice vectors for the array.
For 1-d targets we let $m$ vary, but set $n=0$.
For 2-d targets,
the area per TUC is
\beq
\ATUC \equiv |\bLu\times\bLv| = L_u L_v \sin\theta_{uv}
\eeq
where
$\theta_{uv}$ is the angle
between $\bLu$ and $\bLv$.

The replica dipole polarization $\bP_{jmn}(t)$ 
is phase-shifted relative to 
$\bP_{j00}(t)$:
\beqa 
\bP_{jmn}(t)
&=& \bP_{j00}(t)\exp\left[i(m \bk_0\cdot\bLu + n \bk_0\cdot\bLv)\right]
\eeqa

Define a matrix $\bA$ such that $-\bA_{j,kmn}\bP_{kmn}$ 
gives the electric field $\bE$ at  $\br_{j00}$ 
produced by an oscillating dipole $\bP_{kmn}$
located at $\br_{kmn}$.
Expressions for the 3$\times$3 tensor elements 
of $\bA$ have been presented elsewhere
\cite[e.g.,][]{Draine+Flatau_1994};
$\bA$ depends on the target geometry and wavelength of the incident
radiation, but not on the target composition or on the
direction or polarization state of the incident wave.

Using eq.\ (\ref{eq:P periodicity}) we may construct
a matrix $\bAtilde$ such that, for $j\neq k$,
$-\bAtilde_{j,k}\bP_{k00}$ gives the electric field
at $\br_{j00}$ produced by a dipole $\bP_{k00}$ and {\it all of its replica
dipoles} $\bP_{kmn}$, and for $j=k$ it gives the electric field 
at $\br_{j00}$ produced only by the replica dipoles:
\beqa 
\label{eq:APBCtwod}
\bAtilde_{j,k}\! &=&\! \sum_{m=-\infty}^\infty 
\sum_{n=-n_{\rm max}}^{n_{\rm max}}
\left(1-\delta_{jk}\delta_{m0}\delta_{n0}\right)
\bA_{j,kmn} 
\exp\left[i(m\bk_0\cdot\bLu + n\bk_0\cdot\bLv)\right] ~.~~~
\eeqa
where
$n_{\rm max}=0$ for 1-d targets and $n_{\rm max}=\infty$ for 2-d targets,
and $\delta_{ij}$ is the Kronecker delta.
For $|m|,|n|\rightarrow\infty$, location $j00$ is in the radiation zone
of dipole $kmn$, and the electric field falls off in magnitude
only as $1/r$.  
The sums in (\ref{eq:APBCtwod}) would be divergent were it not
for the oscillating phases of the terms, which ensure convergence.
Evaluation of these sums can be computationally-demanding when
$k_0L_y$ or $k_0L_z$ are small.
Chaumet et al.\ \cite{Chaumet+Rahmani+Bryant_2003} 
have discussed methods for
efficient evaluation of these sums.

We 
evaluate (\ref{eq:APBCtwod}) numerically by introducing a factor
$\exp[-(\gamma k_0r)^4]$ to smoothly suppress the contributions from large
$r$, and truncating the sums: 
\beqa
\bAtilde_{j,k}\!&\approx&\! 
{\sum_{m,n}}^\prime
\bA_{j,kmn} 
\exp\left[i(m\bk_{0}\cdot\bLu + n\bk_{0}\cdot\bLv)-(\gamma k_0r_{j,kmn})^4
\right]
\label{eq:APBC,2d,sum}
~~~
\eeqa
where $r_{j,kmn}\equiv|\br_{kmn}-\br_{j00}|$ and the summation is
over $(m,n)$ with $r_{j,kmn}\leq 2/\gamma k_0$, i.e., out to distances
where the
suppression factor $\exp\left[-(\gamma k_0r)^4\right]\approx e^{-16}$.
For given $\bk_0$, $\bLu$, $\bLv$, the $\bAtilde_{j,k}$
depend only on $\br_{j00}-\br_{k00}$, and therefore only
$O(8N)$ distinct $\bAtilde_{j,k}$ require evaluation.

Ideally, one would use a very small value for the interaction cutoff
parameter $\gamma$, but the number of terms 
[$\propto \gamma^{-1}$ for 1-d, or
$\propto \gamma^{-2}$ for 2-d]
in eq.\ (\ref{eq:APBC,2d,sum})
diverges as $\gamma\rightarrow0$.
We show that $\gamma\approx 0.001$ ensures
accurate results for the cases studied here.

The polarizations $\bP_{j00}$ 
of the dipoles in the TUC must satisfy the
system of equations
\beq \label{eq:SLE}
\bP_{j00} = \balpha_j \left[
\bE_\inc(\br_j) - \sum_{k\neq j} \bAtilde_{j,k}\bP_{k00}
\right]
~.~~
\eeq
If there are $N$ dipoles in one TUC, then (\ref{eq:SLE}) is a system
of $3N$ linear equations
where the polarizability tensors $\balpha_j$ are obtained from
lattice dispersion relation theory
\cite{Draine+Goodman_1993,Gutkowicz-Krusin+Draine_2004}.
After $\bAtilde$ has been calculated, 
equations (\ref{eq:SLE}) can be solved for $\bP_{j00}$ using 
iterative techniques when $N\gg 1$.

\section{\label{sec:in the radiation zone}
          In the Radiation Zone}

In the radiation zone $kr \gg 1$, the electric field due to dipole $jmn$ is
\beq
\bE_{jmn} = \frac{k_0^2 \exp\left(ik_0|\br -\br_{jmn}|\right)}{|\br-\br_{jmn}|}
\left[1 - \frac{(\br-\br_{jmn})(\br-\br_{jmn})}{|\br-\br_{jmn}|^2}\right]
\bP_{jmn}~,~~
\eeq
\beqa
|\br-\br_{jmn}| &=& \left[r^2-2\br\cdot\br_{jmn}+r_{jmn}^2\right]^{1/2}
\\ \nonumber
&\approx&r
\left\{1-\frac{\br\cdot\br_{jmn}}{r^2}
+\frac{1}{2r^2}
\left[r_{jmn}^2-\left(\frac{\br\cdot\br_{jmn}}{r}\right)^2\right] + ...
\right\}
~.~~
\label{eq:|r-rjmn|}
\eeqa
Define the unit vector $\khat_s\equiv \bk_s/k_0$.
We seek to sum the contribution of all the dipoles to the electric field
propagating in direction $\khat_s$.
At location $\br = r\khat_s$, the dominant contribution will be
from dipoles located within the 
Fresnel zone
\cite[see, e.g., ref.][]{Born+Wolf_1999},
which will have a transverse
radius $\RF\approx (r/k_0)^{1/2}$ .
For dipoles within the Fresnel zone,
\beq
\frac{1}{|\br-\br_{jmn}|}
\left[1 - \frac{(\br-\br_{jmn})(\br-\br_{jmn})}{|\br-\br_{jmn}|^2}\right]
\bP_{jmn}
\approx 
\frac{1}{r}\left[1 - \khat_s \khat_s\right]\bP_{jmn}
~~~.
\eeq
Thus, in the radiation zone,
\beq
\bE(\br) = \frac{k_0^3}{k_0r}\exp\left(ik_0r\right)
\left[1-\khat_s\khat_s\right]
\sum_j \bP_{j00} \sum_{m,n} \exp\left(i\Psi_{jmn}\right)
\eeq
\beqa \label{eq:Psi def}
\Psi_{jmn} 
\!&\equiv&\! 
m\bk_{0}\cdot\bLu+n\bk_{0}\cdot\bLv -\bk_s\cdot\br_{jmn} +
\frac{k_0}{2r}\left[r_{jmn}^2 - \left(\khat_s\cdot\br_{jmn}\right)^2\right]
\\
\nonumber
&\approx&\!
             -\bk_s\cdot\br_{j00} +
             m(\bk_0-\bk_{s})\cdot\bLu +
             n(\bk_0-\bk_{s})\cdot\bLv
\\
  &&\!         +\frac{1}{2k_0r}\bigg[ m^2(k_0^2-k_{su}^2)L_u^2
             +n^2(k_0^2-k_{sv}^2)L_v^2
         +2mn(k_0^2\bLu\!\cdot\!\bLv\!-\!k_{su}k_{sv}L_uL_v)
                          \bigg]
             \!+\! O\left(\frac{mL}{r}\right),~~~~~
\eeqa
where 
$k_{su}\equiv \bk_s\cdot\bLu/L_u$,
$k_{sv}\equiv \bk_s\cdot\bLv/L_v$,
and terms of order $(mL/r)$ may be neglected because
$mL/r \sim \RF/r \propto r^{-1/2}$ as $r\rightarrow \infty$.
Thus, for $r\rightarrow\infty$, the electric field produced by the
oscillating dipoles is
\beq \label{eq:EFG}
\bE_s = \left\{\frac{k_0^2}{r}\exp\left(ik_0r\right)
\left[1-\khat_s\khat_s\right]
\sum_j \bP_{j00}\exp\left(-i\bk_s\cdot\br_{j00}\right)\right\} G(r,\bk_s)
\eeq
\beq \label{eq:Gsum}
G(r,\bk_s) \equiv \sum_{m,n} \exp\left(i\Phi_{mn}\right)
\eeq
\beqa
\nonumber
\Phi_{mn} &\equiv& 
m(\bk_{0}-\bk_{s})\cdot \bLu + 
n(\bk_{0}-\bk_{s})\cdot \bLv 
\\
&&
+\frac{1}{2k_0r}
 \left[m^2(k_0^2-k_{su}^2)L_u^2
     +n^2(k_0^2-k_{sv}^2)L_v^2
     +2mn(k_0^2\bLu\cdot\bLv-k_{su}k_{sv}L_uL_v)
                          \right]
\label{eq:Phi_mn}
~.~~
\eeqa
It is convenient to define
\beq
\FTUC(\khat_s) \equiv k_0^3
\left[1-\khat_s\khat_s\right]\sum_{j=1}^N \bP_{j00}
\exp\left(i\omega t-i\bk_s\cdot\br_{j00}\right)
~,~~
\eeq
so that the electric field produced by the dipoles is
\beq
\bE_s = \frac{\exp\left(i\bk_s\cdot\br-i\omega t\right)}{k_0r}
\FTUC(\khat_s) G(r,\bk_s)
~~~.
\eeq
$\FTUC$ depends upon the scattering direction $\khat_s$, and
also upon the direction of incidence $\khat_0$ and polarization 
$\bE_0$ of the incident wave.

\subsection{Isolated Finite Target: $\ndim=0$}

We will refer to finite isolated targets --
consisting of only the dipoles in a single TUC --
as targets that are periodic in 
$\ndim=0$ dimensions.
For this case, we simply set $G=1$ in eq.\ (\ref{eq:EFG}); 
the scattered electric field in the radiation zone is
\beq
\bE_s=\frac{\exp\left(ik_0r-i\omega t\right)}{k_0r}\FTUC(\khat_s)
~~~.
\eeq
The time-averaged scattered intensity is
\beqa
I_s = \frac{c|\bE_s|^2}{8\pi} 
= \frac{c}{8\pi k_0^2r^2} |\FTUC|^2
~~~,
\eeqa
and the differential scattering cross section is
\beq
\frac{dC_{\rm sca}}{d\Omega} = \frac{1}{k_0^2} 
\frac{|\bF_{\rm TUC}|^2}{|\bE_0|^2}
~~~.
\eeq

\subsection{Target Periodic in One Dimension: $\ndim=1$}

Without loss of generality, we may assume that targets with
1-d periodicity
repeat in the $\yhat$ direction.
It is easy to see from eq.(\ref{eq:Gsum}) and (\ref{eq:Phi_mn})
that $G=0$ except 
for scattering directions satisfying
\beq \label{eq:integer M}
k_{sy} = k_{0y} + M\frac{2\pi}{L_y}~~~ , ~~~M=0,\pm 1, \pm2, ...,
\eeq
where energy conservation ($k_s^2=k_0^2$) 
limits the allowed values of the integer $M$:
\beq \label{eq:energy conservation}
(-k_{0y}-k_0)\frac{L_y}{2\pi} \leq M \leq (-k_{0y}+k_0)\frac{L_y}{2\pi}
~~~.
\eeq
If $(k_0+|k_{0y}|)L_y<2\pi$, then only $M=0$ scattering is allowed.
Define polar angles $\alpha_0$ and $\alpha_s$
for the incident and scattered radiation, so that
\beqa
k_{0y} &=& k_0\cos\alpha_0
\\
k_{sy} &=& k_0\cos\alpha_s
~~~.
\eeqa
For each allowed value of $k_{sy}$, the scattering directions define a cone:
\beqa 
\bk_s &=& k_{sy}\yhat + 
(k_0^2-k_{sy}^2)^{1/2}
\frac{\sin\alpha_s}{\sin\alpha_0}
\times\left[
(\hat{\bf k}_0-\yhat \cos\alpha_0)\cos\polang + \yhat\times\khat_0\sin\polang
\right]~,~
\label{eq:ks for 1d}
\eeqa
where $\polang$ is an azimuthal angle measured around the target axis $\yhat$.
The sum for $G(r,\bk_s)$ is [since $M$ must be an integer -- 
see eq.\ (\ref{eq:integer M})] 
\beqa
\!\!G\!\! = \!\!\sum_{m=-\infty}^\infty \exp\left(i\Phi_{m0}\right) 
&=& \!\!
\sum_{m=-\infty}^\infty
\exp\left[-2\pi i M m + i \frac{m^2}{2k_0r}(k_0^2-k_{sy}^2)L_y^2
\right]~~~~~
\\
~~~~~&\rightarrow& 
\lim_{\epsilon\rightarrow0^+}
\int_{-\infty}^{\infty} dm 
\exp\!\!\left[\frac{i(1+i\epsilon)}{2k_0r}m^2(k_0^2-k_{sy}^2)L_y^2\right]~~~~~~
\\
&=&
\frac{(2\pi i k_0r)^{1/2}}{(k_0^2-k_{sy}^2)^{1/2}L_y}
=\frac{(2\pi i k_0r)^{1/2}}{k_0 L_y\sin\alpha_s}
~~~,
\eeqa
and the scattered electric field
\beq
\bE_s = \left(\frac{2\pi i}{k_0r}\right)^{1/2}
\frac{\exp\left(ik_0r-i\omega t\right)}{k_0L_y \sin\alpha_s}
\bF_{\rm TUC}(\khat_s)
\eeq
shows the expected $r^{-1/2}$ behavior far from the scatterer
(the distance from the cylinder axis is $R=r\sin\alpha_s$).

For each allowed value of $\hat{\bf k}_{s}$, 
the total time-averaged scattered power $\Pbar_{\rm sca}$,
per unit length of the target, per unit azimuthal angle $\polang$, may be
written
\beq
\frac{d^2\Pbar_{\rm sca}}{dLd\polang} =
\frac{|\bE_0|^2}{8\pi} c\,\frac{d^2\Csca}{dLd\polang}
~~~,
\eeq
where the differential scattering cross section is
\beqa
\frac{d^2C_{\rm sca}}{dLd\polang}
&=&
\frac{8\pi}{ |\bE_0|^2}
\frac{1}{c}
\frac{d^2\Pbar_{\rm sca}}{dLd\polang}
=
\frac{8\pi}{ |\bE_0|^2 c} 
\frac{|\bE|^2 c}{8\pi} R\sin\alpha_s
\\
&=&
\frac{2\pi}{k_0^3L_y^2}
\frac{\left|\bF_{\rm TUC}\right|^2}{|\bE_0|^2}
~~~.
\eeqa

\subsection{Target Periodic in Two Dimensions: $\ndim=2$}

For targets that are periodic in two dimensions,
it is apparent from eq.\ (\ref{eq:Gsum}) and (\ref{eq:Phi_mn})
that $G=0$ unless 
\beqa 
\label{eq:2d condition1}
\left(\bk_{s}-\bk_0\right)\cdot\bLu &=& {2\pi}M~~ , ~~M=0,\pm 1, \pm2, ...,
\\
\label{eq:2d condition2}
\left(\bk_{s}-\bk_0\right)\cdot\bLv &=& {2\pi}N~~ , ~~N=0,\pm 1, \pm2, ... .~~~
\eeqa
The 2-d target constitutes 
a diffraction grating, with scattering allowed only in
directions given by (\ref{eq:2d condition1},\ref{eq:2d condition2}).
It is convenient to define the reciprocal lattice vectors
\beq
\bu \equiv \frac{2\pi \xhat \times \bLv}{\xhat\cdot(\bLu\times\bLv)}
~~~,~~~
\bv\equiv\frac{2\pi\xhat\times\bLu}{\xhat\cdot(\bLv\times\bLu)}
\eeq
The wave vector transverse to the surface normal is
\beq \label{eq:ksperp}
\bk_{s\perp}\equiv \bk_{0\perp}+M\bu+N\bv
~~.
\eeq
Energy conservation requires that
\beq 
\nonumber
k_{sx}^2 
\label{eq:MN condition}
=
k_0^2 - |\bk_{0\perp}+M\bu+N\bv|^2 > 0
~.~~~~
\eeq
For any $(M,N)$ allowed by eq.\ (\ref{eq:MN condition}), there
are two allowed values of $k_{sx}$, differing by a sign; one 
(with $k_{sx}k_{0x}>0$)
corresponds
to the $(M,N)$ component of the transmitted wave, and the other
(with $k_{sx}k_{0x}<0$)
to the $(M,N)$ component of the reflected wave.
Define
\beqa
\sin\alpha_0 &\equiv& \frac{|k_{0x}|}{k_0} ~~~,
\\
\sin\alpha_s &\equiv& \frac{|k_{sx}|}{k_0} ~~~.
\eeqa
Note that $\alpha_0=\pi/2$ for normal incidence, and $\alpha_0\rightarrow 0$
for grazing incidence.
For $\bk_s$ satisfying eq. (\ref{eq:2d condition1},\ref{eq:2d condition2}), 
we have
\beqa 
G &=& \sum_{m,n} \exp\left(i\Phi_{mn}\right) \nonumber
\\
&=& 
\sum_{m,n}
\exp\left\{
     \frac{i}{2k_0r}
     \left[\left(k_0^2-k_{su}^2\right)L_u^2m^2 
     \right.\right.
        + \left(k_0^2-k_{sv}^2\right)L_v^2n^2
        +  2\left(k_0^2\bLu\cdot\bLv -k_{su}k_{sv}L_uL_v\right)
           mn\Big]\!\bigg\}~~~
     \nonumber
\\
&\rightarrow& \lim_{\epsilon\rightarrow0^+}
\int_{-\infty}^{\infty} dm \int_{-\infty}^\infty dn    \nonumber
\\&&
\times\exp\bigg\{\frac{i(1+i\epsilon)}{2k_0r}
\Big[
(k_0^2-k_{su}^2)L_u^2 m^2
+ 
k_0^2-k_{sv}^2)L_v^2 n^2 
  +2(k_0^2\bLu\cdot\bLv\!-\!k_{su}k_{sv}L_u L_v) mn\Big]\!\bigg\} \nonumber
\\
\nonumber
&=&\lim_{\epsilon\rightarrow0^+}\frac{1}{\ATUC}
\int_{-\infty}^\infty dy \int_{-\infty}^\infty dz
\exp
\bigg\{
\frac{i(1\!+\!i\epsilon)}{2k_0r}
\Big[
k_0^2(y^2\!+\!z^2)-(k_{sy}y+k_{sz}z)^2
\Big]
\bigg\}
\\
&=& 
\frac{2\pi i r}{k_0 \ATUC\sin\alpha_s}
~.~~~~
\eeqa
The scattered electric field is
\beq
\bE_s = 
\frac{2\pi i \exp\left(i\bk_s\cdot\br-i\omega t\right)}{k_0^2\ATUC
\sin\alpha_s}
\bF_{\rm TUC}(\khat_s)
~~~.
\eeq
Note that $|\bE_s|$ 
is independent of distance $x=r\sin\alpha_s$ from the target,
as expected for a target that is infinite in two directions.
For pure forward scattering, $\bk_s=\bk_0$, we must sum the
incident wave $\bE_\inc$ and the radiated wave $\bE_s$:
\beq \label{eq:2d forward E}
\bE = \exp\left(i\bk_0\cdot\br-i\omega t\right) 
\left[\bE_0+ 
\frac{2\pi i \,
\bF_{\rm TUC}(\khat_s\!=\!\khat_0)}
{k_0^2\ATUC\sin\alpha_s}
\right]
~.~~
\eeq
The cross section per unit target area $A$ for scattering into direction
$(M,N)$ is (for $\bk_s\neq\bk_0$):
\beqa
\frac{d\Csca(M,N)}{dA} &=& \frac{|\bE|^2\sin\alpha_s}{|\bE_0|^2\sin\alpha_0}
\\
&=&
\frac{4\pi^2}{k_0^4\ATUC^2\sin\alpha_0\sin\alpha_s}
\frac{\left|\bF_{\rm TUC}(\khat_s)\right|^2}{|\bE_0|^2}
~~~,
\eeqa
where $\Csca$ can be evaluated for either transmitted or reflected waves.
For the special case $\bk_s=\bk_0$, the transmission coefficient $T(M,N)$
is obtained from the total forward-propagating wave (\ref{eq:2d forward E}):
\beq
T(0,0) = \frac{1}{|\bE_0|^2}
\left| \bE_0 + 
      \frac{2\pi i~
\bF_{\rm TUC}(\khat_s\!=\!\khat_0)
}{k_0^2\ATUC\sin\alpha_0}\right|^2
~~~.
\eeq

\section{\label{sec:S_i}
          Scattering Amplitude Matrices $S_i^{(\ndim d)}$}

\subsection{Isolated Finite Targets: $\ndim=0$}

In the radiation zone, 
the scattered electric field is related to the incident electric field
via the $2\times2$ scattering amplitude matrix
\cite{Bohren+Huffman_1983}, 
defined so that
\beqa
\left(
\begin{array}{c}
	{\bf E}_s\cdot\ehat_{s\parallel}\\
	{\bf E}_s\cdot\ehat_{s\perp}
\end{array}
\right)
&=&
{i\exp(i{\bk_s}\cdot{\br}-i\omega t)\over k_0r}
\left(
\begin{array}{cc}
        S_2^{(0d)}&S_3^{(0d)}\\
	S_4^{(0d)}&S_1^{(0d)}
\end{array}
\right)
\left(
\begin{array}{c}
	{\bf E}_0\cdot\ehat_{i\parallel}\\
	{\bf E}_0\cdot\ehat_{i\perp}
\end{array}
\right) ~,~~~
\label{eq:f_ml_def}
\eeqa
where
\beqa
\ehat_{i\perp} = \ehat_{s\perp}
&\equiv&
\frac{\khat_s\times\khat_0}{|\khat_s\times\khat_0|}
=
\frac{\khat_s\times\khat_0}{1-(\khat_s\cdot\khat_0)^2}
=
-\phihat_s
\\ \label{eq:behat_iparallel}
\ehat_{i\parallel}
&\equiv&
\khat_0\times\ehat_{i\perp}
=
\frac{\khat_s-(\khat_s\cdot\khat_0)\khat_0}
{1-(\khat_s\cdot\khat_0)^2}
\\ \label{eq:behat_sparallel}
\ehat_{s\parallel}
&\equiv&\khat_s\times\ehat_{s\perp}
=
\frac{-\khat_0 + (\khat_s\cdot\khat_0)\khat_s}
{1-(\khat_s\cdot\khat_0)^2}
=\thetahat_s ~~~~
\eeqa
are the usual conventions for the incident and scattered
polarization vectors parallel and
perpendicular to the scattering plane 
\cite[see, e.g., \S3.2 of ref.][]{Bohren+Huffman_1983}.

\subsection{Target Periodic in One Dimension: $\ndim=1$}

For targets with 1-d periodicity,
it is natural to generalize the scattering amplitude
matrix, so that -- for directions $\khat_s$ 
for which scattering is allowed --
the scattered electric field at a distance $R=r\sin\alpha_s$ from the target
is
\beqa
\left(
\begin{array}{c}
	{\bf E}_s\cdot\ehat_{s\parallel}\\
	{\bf E}_s\cdot\ehat_{s\perp}
\end{array}
\right)
&=&
{i\exp(i{\bk_s}\cdot\br-i\omega t)\over (k_0R)^{1/2}}
\left(
\begin{array}{cc}
	S_2^{(1d)}&S_3^{(1d)}\\
	S_4^{(1d)}&S_1^{(1d)}
\end{array}
\right)
\left(
\begin{array}{c}
	{\bf E}_0\cdot\ehat_{i\parallel}\\
	{\bf E}_0\cdot\ehat_{i\perp}
\end{array}
\right) ~~~~
\label{eq:S for 1d periodicity}
\label{eq:S_j(1d)_def}
\eeqa
for $\bk_s$ satisfying eq.\ (\ref{eq:integer M}-\ref{eq:ks for 1d}).

\subsection{Target Periodic in Two Dimensions: $\ndim=2$}

For targets with 2-d periodicity,
it is natural to generalize the scattering amplitude matrix so that,
for directions $\bk_s\neq\khat_0$ for which scattering is allowed, we write
\beqa
\left(
\begin{array}{c}
	{\bf E}_s\cdot\ehat_{s\parallel}\\
	{\bf E}_s\cdot\ehat_{s\perp}
\end{array}
\right)
&=&
i\exp(i\bk_s\cdot\br-i\omega t)
\left(
\begin{array}{cc}
	S_2^{(2d)}&S_3^{(2d)}\\
	S_4^{(2d)}&S_1^{(2d)}
\end{array}
\right)
\left(
\begin{array}{c}
	{\bf E}_0\cdot\ehat_{i\parallel}\\
	{\bf E}_0\cdot\ehat_{i\perp}
\end{array}
\right)~~~~
\label{eq:S_j(2d)_def}
\eeqa
for $\bk_s$ satisfying eq.\ (\ref{eq:ksperp}-\ref{eq:MN condition}).

For the special case
of forward scattering ($M=N=0$ and $\bk_s=\bk_0$), where the scattering
plane is not simply defined by $\bk_0$ and $\bk_s$, it is natural to
use $\bk_0$ and the target normal $\xhat$ to define the scattering plane.
Thus
\beqa
\ehat_{i\perp} = \ehat_{s\perp} &\equiv& 
\frac{\bk_0\times\bk_{s\perp}}{|\bk_0\times\bk_{s\perp}|}
\eeqa
with $\ehat_{i\parallel}$ and $\ehat_{s\parallel}$ defined by
(\ref{eq:behat_iparallel},\ref{eq:behat_sparallel}).
For $r\rightarrow\infty$
\beqa
\left(
\begin{array}{c}
	{\bf E}\cdot\ehat_{s\parallel}\\
	{\bf E}\cdot\ehat_{s\perp}
\end{array}
\right)
&=&
i\exp(i\bk_0\cdot{\bf r}-i\omega t)
\!\!\left(
\begin{array}{cc}
	(S_{2}^{(2d)}\!-\!i)&0\\
	0&(S_{1}^{(2d)}\!-\!i)
\end{array}
\right)
\!
\left(
\begin{array}{c}
	{\bf E}_0\cdot\ehat_{i\parallel}\\
	{\bf E}_0\cdot\ehat_{i\perp}
\end{array}
\right).~~~
\label{eq:MN=00 forward scattering}
\eeqa

\subsection{Far-Field Scattering Amplitude Matrices}

The scattering amplitude matrices $S_i^{(\ndim d)}(\bk_s)$
are directly related to the
$\FTUC$ for the three cases, $\ndim=0,1,2$:
\beqa
S_1^{(\ndim d)} &=& C_\ndim\ehat_{s\perp}\cdot
               \FTUC(\khat_s,\bE_0=\ehat_{i\perp})
\\
S_2^{(\ndim d)} &=& C_\ndim\ehat_{s\parallel}\cdot 
               \FTUC(\khat_s,\bE_0=\ehat_{i\parallel})
\\
S_3^{(\ndim d)} &=& C_\ndim\ehat_{s\parallel}\cdot
               \FTUC(\khat_s,\bE_0=\ehat_{i\perp})
\\
S_4^{(\ndim d)} &=& C_\ndim\ehat_{s\perp}\cdot 
                \FTUC(\khat_s,\bE_0=\ehat_{i\parallel})
\eeqa
\beqa
C_0 &=& -i
\\
C_1 &=& -\left(\frac{2\pi i}{\sin\alpha_s}\right)^{1/2}\frac{i}{k_0L_y}
\\
C_2 &=& \frac{2\pi}{k_0^2\ATUC\sin\alpha_s}
\eeqa
for finite targets ($C_0$), and targets that are periodic in one 
or two dimensions ($C_1$ or $C_2$).

\section{\label{sec:mueller matrix}
         Far-Field Scattering Matrix for Stokes Vectors}

For a given scattering direction $\khat_s$, the $2\times2$ complex
amplitude matrix $S_i^{(\ndim d)}(\bk_s)$ fully characterizes the far-field
scattering properties of the target.
The far-field scattering properties of 
an isolated finite target 
are characterized
by the $4\times4$ dimensionless Mueller matrix $S_{\alpha\beta}^{(0d)}$,
with the Stokes vector of radiation scattered into direction $\khat_s$
at a distance $r$ from the target given by
\beq
I_{{\rm sca},\alpha} 
\equiv \frac{1}{(k_0r)^2} 
\sum_{\beta=1}^4 S_{\alpha\beta}^{(0d)} I_{{\rm inc},\beta}
~~~,
\eeq
where $I_{{\rm inc},\beta}=(I,Q,U,V)_{\rm inc}$ 
is the Stokes vector for the radiation incident on the
target.
For 1-d targets, we define the dimensionless
scattering matrix
$S_{\alpha\beta}^{(1d)}$ by
\beq
I_{{\rm sca},\alpha} 
\equiv \frac{1}{k_0R} 
\sum_{\beta=1}^4 S_{\alpha\beta}^{(1d)} I_{{\rm inc},\beta} 
~~~,
\eeq
where $R$ is the distance from the one-dimensional target.
For 2-d targets, we define
$S_{\alpha\beta}^{(2d)}$ by
\beq
I_{{\rm sca},\alpha} \equiv\sum_{\beta=1}^4 S_{\alpha\beta}^{(2d)} 
I_{{\rm inc},\beta} 
~~~.
\eeq
The $4\times4$ scattering intensity matrix $S_{\alpha\beta}^{(\ndim d)}$ 
is obtained from the scattering amplitude matrix elements $S_{i}^{(\ndim d)}$.
Except for the special case of forward scattering ($\bk_s=\bk_0$) for
2-d targets, the equations are the same as
eq.\ (3.16) of 
Bohren and Huffman \cite{Bohren+Huffman_1983}.
For example
\beqa
S_{11}^{(\ndim d)}\!\! &=& \!\!\frac{1}{2}\left(
                 |S_{1}^{(\ndim d)}|^2 +
                 |S_{2}^{(\ndim d)}|^2 +
                 |S_{3}^{(\ndim d)}|^2 +
                 |S_{4}^{(\ndim d)}|^2
                 \right) ~~~~
\\
S_{21}^{(\ndim d)}\!\! &=& \!\!\frac{1}{2}\left(
                 |S_{2}^{(\ndim d)}|^2 -
                 |S_{1}^{(\ndim d)}|^2 -
                 |S_{4}^{(\ndim d)}|^2 +
                 |S_{3}^{(\ndim d)}|^2
                 \right) ~~~~~
\\
S_{14}^{(\ndim d)}\!\! &=&\!\! {\rm Im}\left(
                  S_{2}^{(\ndim d)}S_{3}^{(\ndim d)*}
                 -S_{1}^{(\ndim d)}S_{4}^{(\ndim d)*}
		 \right) ~.~~~
\eeqa
For the special case of forward scattering ($\khat_s=\khat_0$)
for 2-d targets, it is necessary to replace $S_1^{(2d)}$ and
$S_2^{(2d)}$ with $(S_1^{(2d)}-i)$ and $(S_2^{(2d))}-i)$ 
[cf. Eq. (\ref{eq:MN=00 forward scattering})]. Thus, for example,
\beq
S_{11}^{(2d)}(\bk_s\!=\!\bk_0) = 
\frac{1}{2}\left(
                 |S_1^{(2d)}\!-\!i|^2 +
                 |S_2^{(2d)}\!-\!i|^2 +
                 |S_3^{(2d)}|^2 +
                 |S_4^{(2d)}|^2
                 \right)
                 .~~
\eeq

\begin{figure}[t]
  \begin{center}
  \includegraphics[angle=0,width=5.75in]{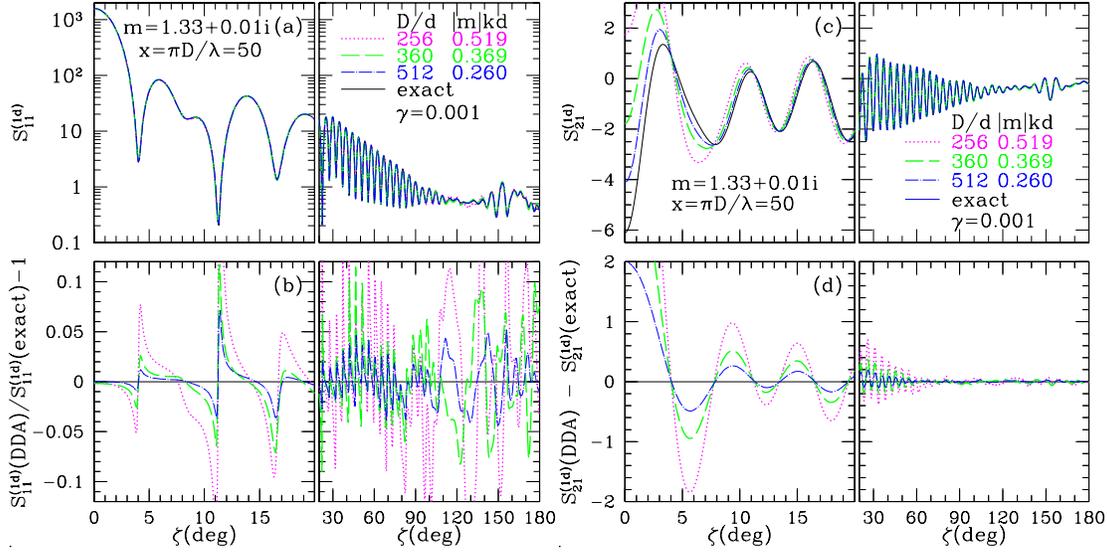}   
  \caption{\label{fig:S11,S21_50} 
     Scattering by an infinite cylinder with diameter $D$ and
     $m=1.33+0.01i$, for radiation with $x=\pi D/\lambda = 50$ and
     incidence angle $\alpha_0=60^\circ$.
     (a) $S_{11}^{(1d)}$.
     Solid curve: exact solution.
     Broken curves: DDA results for $D/d=$ 256, 360, and 512
     ($N=51676$, 102036, 206300 dipoles
     per TUC);
     (b) fractional error in $S_{11}^{(1d)}({\rm DDA})$.
     (c) $S_{21}^{(1d)}$; (d) error in $S_{21}^{(1d)}$.}
   \end{center}
\end{figure}

\begin{figure}[t]
  \begin{center}
  \includegraphics[angle=0,width=5.75in]{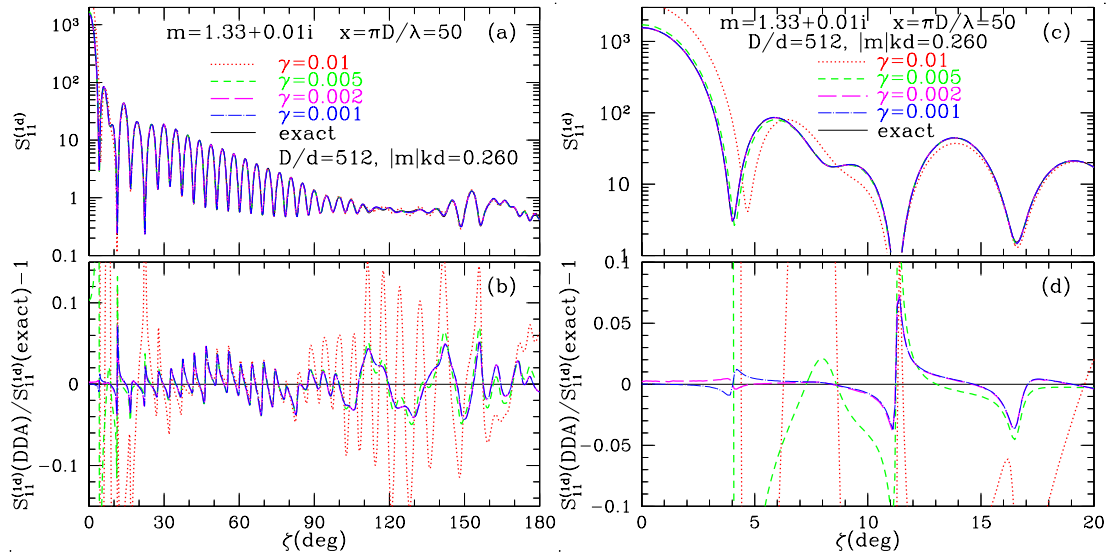}
  \caption{\label{fig:S11_50_gamma} 
     Scattering by an infinite cylinder with diameter $D$
     and
     $m=1.33+0.01i$, for radiation with $\pi D/\lambda = 50$,
     and incidence angle $\alpha_0=60^\circ$.
     (a) Exact solution (solid curve) and 
     DDA results for $D/d=512$ and
     various values of the interaction cutoff parameter $\gamma$;
     (b) fractional error in $S_{11}^{(1d)}$;
     (c,d) same as (a,b), but expanding the region $0<\zeta<20^\circ$.
     For this case, results computed with $\gamma=0.002$ and $0.001$
     are nearly indistinguishable.
     }
   \end{center}
\end{figure}

\section{\label{sec:T and R}
         Transmission and Reflection Coefficients for 2-D Targets}

For targets with 2-d periodicity, 
it is natural to define generalized 
transmission and reflection coefficients for the Stokes vectors:
for scattering order $(M,N)$,
$I_{{\rm sca},\alpha} = \sum_\beta T_{\alpha\beta}(M,N)I_{{\rm inc},\beta}$
is the 
Stokes vector component $\alpha$ for radiation 
with $k_{sx}k_{{\rm inc},x}>0$, and
$R_{\alpha\beta}(M,N)$ 
is the fraction of the incident Stokes vector component $\beta$
that emerges in Stokes vector component $\alpha$
with $k_{sx}k_{{\rm inc},x}<0$.
These can be related to the $S_{\alpha\beta}^{(2d)}$:
\beqa \label{eq:R_ab}
R_{\alpha\beta}(M,N) &=&
\frac{\sin\alpha_s}{\sin\alpha_0} S_{\alpha\beta}^{(2d)}
~~~{\rm for~} k_{sx}k_{0x} < 0
~~~,
\\ \label{eq:T_ab}
T_{\alpha\beta}(M,N) &=& 
\frac{\sin\alpha_s}{\sin\alpha_0} S_{\alpha\beta}^{(2d)}
~~~{\rm for~} k_{sx}k_{0x} > 0
~~~,
\eeqa
The fraction of the incident power that is absorbed by the target
is
\beq
\frac{P_{\rm abs}/{\rm Area}}{|\bE_0|^2 c \sin\alpha_0/8\pi}
=1-\sum_{M,N}\sum_{\beta=1}^4
\left[R_{1\beta}(M,N)+T_{1\beta}(M,N)\right]
\frac{I_{\inc,\beta}}{I_{\inc,1}} ~~~,
\eeq
where $I_{\inc,\beta}$ is the Stokes vector of the incident radiation.

For unpolarized incident radiation, $R_{11}(M,N)$ is the fraction of the
incident power that is reflected in diffraction component $(M,N)$,
$T_{11}(M,N)$ is the fraction that is transmitted in component $(M,N)$,
and $1-\sum_{M,N}[R_{11}(M,N)+T_{11}(M,N)]$ is the fraction of the incident
power that is absorbed.

\begin{figure}[t]
  \begin{center}
  \includegraphics[angle=0,width=5.0in]{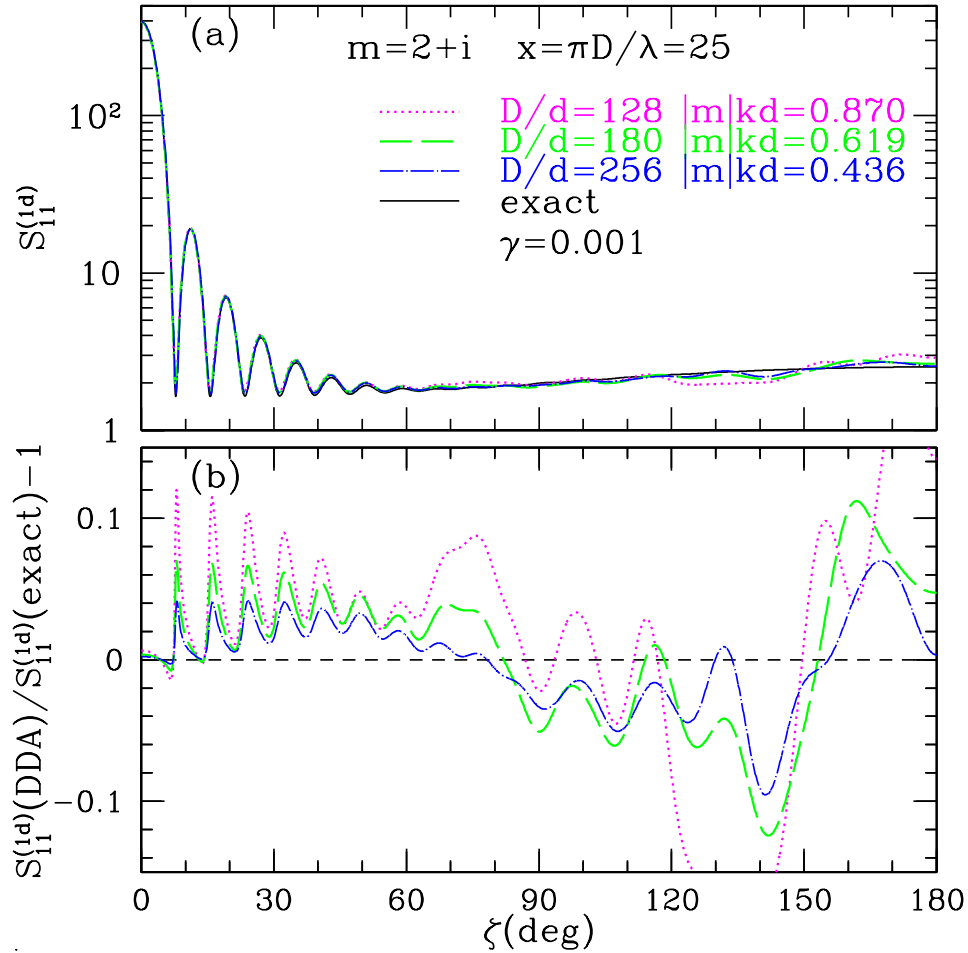}
  \caption{\label{fig:S11b_25} 
     Light scattered by an infinite cylinder with
     $m=2+i$, for radiation with $x=2\pi R/\lambda = 25$ and
     incidence angle $\alpha_0=60^\circ$.
     (a) $S_{11}^{(1d)}$.
     Solid curve: exact solution.
     Broken curves: DDA results for $D/d=$ 128, 180, and 256
     ($N=12972$, 25600, 51676 dipoles
     per TUC).
     (b) Fractional error in $S_{11}^{(1d)}$.
     }
   \end{center}
\end{figure}

\section{\label{sec:cylinder}
          Example: Infinite Cylinder}

DDSCAT 7 has been used to calculate
scattering and absorption by an infinite cylinder consisting of
a periodic array of disks of thickness
$d$ and period $L_y=d$ (where $d$ is the interdipole spacing).
Fig.\ \ref{fig:S11,S21_50}a shows $S_{11}^{(1d)}$
for refractive index $m=1.33+0.01i$ and $\pi D/\lambda=50$
($D$ is the cylinder diameter and $\lambda$ the wavelength of the
incident radiation), and incidence angle $\alpha_0=60^\circ$.
Because $k_0(1+|\cos\alpha_0|)d < 2\pi$, 
equations (\ref{eq:integer M}, \ref{eq:energy conservation}) allow
only $M=0$ scattering, with $\alpha_s=\alpha_0$.
Also shown is the exact solution,
calculated using a code written by 
D. Mackowski (private communication).
Light scattering by cylinders is generally described by scattering
amplitudes $T_i$; in Appendix \ref{app:S_i vs T_i} we provide expressions
relating these $T_i$ to the $S_i$ used here. 
Fig.\  \ref{fig:S11,S21_50}b 
shows the fractional error in $S_{11}^{(1d)}$ calculated
using DDSCAT.  As $d$ 
is decreased, the errors decrease.
Excellent 
accuracy is obtained when the validity criterion
\cite{Draine+Flatau_1994}
$|m|kd\ltsim 0.5$ is satisfied: the fractional error in $S_{11}$
is typically less
than a few \%, except near deep minima in $S_{11}$.

Fig.\ \ref{fig:S11,S21_50}c
shows $S_{21}^{(1d)}$, characterizing scattering of
unpolarized light into the Stokes parameter $Q$ ($S_{21}<0$ corresponds
to linear polarization perpendicular to the scattering plane).
DDSCAT 7 and the exact
solution are in very good agreement when $|m|kd \ltsim 0.5$.
Note that although the error in $S_{21}^{(1d)}(\theta=0)\approx 2$ is
large compared to $S_{21}(0)=-6$, this
is small compared to $S_{11}^{(1d)}(0)\approx 1500$: the scattered radiation
is only slightly polarized.

The results in Fig.\ \ref{fig:S11,S21_50} were obtained
using $\gamma=0.001$ to truncate the integrations.  To see how the results
depend on $\gamma$, Figure \ref{fig:S11_50_gamma} shows $S_{11}^{(1d)}$
computed for the problem of Fig.\ \ref{fig:S11,S21_50} 
but using different values of $\gamma$.
For azimuthal angles $\zeta > 20^\circ$, the results for $\gamma=0.005$
and $0.001$ are nearly indistinguishable; 
the difference between the computed result
and the exact solution is evidently due to the finite number of dipoles used,
rather than the choice of cutoff parameter $\gamma$.  
However, the results for forward scattering are more sensitive to the
choice of $\gamma$, as is seen in Fig.\ \ref{fig:S11,S21_50}c,d:
it is necessary to reduce $\gamma$ to $0.001$ to attain
high accuracy in the forward scattering directions.

Table \ref{tab:table1} gives the CPU times to calculate $\bAtilde$,
to then iteratively solve the scattering problem to
a fractional error $<10^{-5}$ (using double-precision arithmetic), and finally to evaluate the
scattering intensities,
for several of the cases shown in Figs. \ref{fig:S11,S21_50} and
\ref{fig:S11_50_gamma}. 
For most cases the CPU time is dominated by
the iterative solution using the conjugate gradient algorithm.
While the time required to evaluate $\bAtilde$ might be reduced using
the strategies suggested by 
\cite{Chaumet+Rahmani+Bryant_2003}, 
this step
is generally a subdominant part of the computation for targets with
$kd\gtsim 0.1$.

The above results have been for a weakly-absorbing cylinder.  To confirm that
the DDA can be applied to strongly-absorbing material, Fig.\ \ref{fig:S11b_25}
shows scattering calculated for a cylinder with $m=2+i$ 
and $x=\pi D/\lambda=25$.
Once again, the accuracy is very good, with small fractional errors
provided $|m|kd\ltsim 0.5$.

\begin{table}[t]
\caption{\label{tab:table1}
          CPU time to calculate scattering by $m=1.33+0.01i$ 
	  infinite cylinders 
          on single-core 2.4 GHz AMD Opteron model 250}
\begin{ruledtabular}
\begin{tabular}{c r c l l l l}
$\pi D/\lambda$&$N$&$\gamma$&calc.\ $\bAtilde$&solution&scat.&Total\\
               &   &        &(min)            &(min)   &(min) & (min)\\
\hline
25    &  51676 & 0.005      & \ 3.29   & \ 17.6   & 0.65 & \ 22.2 \\
25    &  51676 & 0.001      &  16.2    & \ 17.8   & 0.65 & \ 35.3 \\
50    & 102036 & 0.005      & \ 4.58   & \ 59.7   & 1.27 & \ 66.8 \\
50    & 102036 & 0.001      &  22.8    & \ 45.2   & 1.09 & \ 70.2 \\
50    & 206300 & 0.005      &  13.2    &  254.    & 2.88 &  273.  \\
50    & 206300 & 0.001      &  66.0.   &  292.    & 2.76 &  364.  \\
\end{tabular}
\end{ruledtabular}
\end{table}

\section{\label{sec:slab}
         Example: Plane-Parallel Slab}

\begin{figure}[t]
  \begin{center}
   \includegraphics[angle=0,width=5.75in]{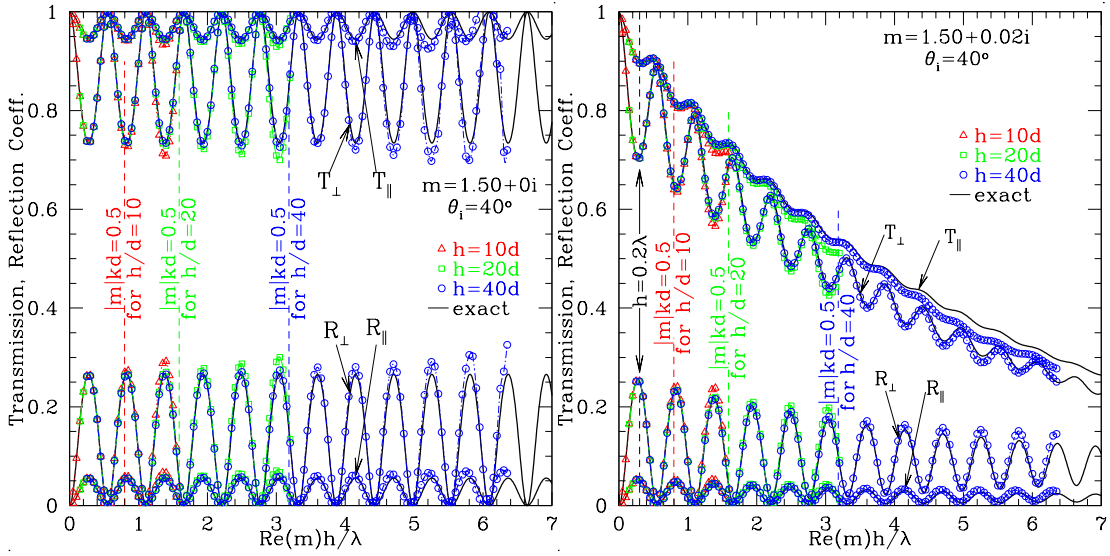}
  \caption{\label{fig:dielectric slab} 
     Transmission and reflection coefficients for radiation with
     wavelength $\lambda$ incident
     at angle $\theta_i=(\pi/2-\alpha_0)=40^\circ$ relative to the normal 
     on a slab with thickness $h$, 
     incident $\bE\parallel$ and $\perp$ to the
     scattering plane, as a function of ${\rm Re}(m)h/\lambda$.
     (a) Nonabsorbing slab with $m=1.5$
     (b) Absorbing slab with $m=1.5+0.02i$.
     Solid curve: exact solution.
     Symbols: results calculated with the DDA using dipole spacing
     $d=h/10$, $h/20$, and $h/40$.
     }
   \end{center}
\end{figure}

Consider a homogeneous plane-parallel slab with thickness $h$ and
refractive index $m$.  Radiation incident on it at angle of incidence
$\theta_i$ will either be specularly reflected or transmitted.
The reflection and transmission coefficients $R$ and $T$ can be calculated
analytically, taking into account multiple reflections within the
slab 
\cite{Born+Wolf_1999}.
With an exact solution in hand, we can evaluate the accuracy of
the DDA applied to this problem.
Figure \ref{fig:dielectric slab} shows results for two cases:
a dielectric slab with $m=1.50$, and an absorbing slab, with
$m=1.50+0.02i$.

DDSCAT 7 was used to calculate reflection, transmission, and absorption
by an infinite slab, generated from a TUC consisting of a single line
of dipoles extending in the $x$ direction, with $L_y=L_z=d$.
The selection rules 
(\ref{eq:2d condition1},\ref{eq:2d condition2},\ref{eq:MN condition})
allow only $M=N=0$: transmission or specular reflection.
The reflection and transmission coefficients for radiation polarized
parallel or perpendicular to the plane containing $\khat$ and the surface
normal are
\beqa
R_\parallel &=& S_{11}^{(1d)}(k_{sx}=-k_{0x})+S_{12}^{(1d)}(k_{sx}=k_{0x})
\\
R_\perp  &=& S_{11}^{(1d)}(k_{sx}=-k_{0x})-S_{12}^{(1d)}(k_{sx}=k_{0x})
\\
T_\parallel &=& S_{11}^{(1d)}(k_{sx}=k_{0x})+S_{12}^{(1d)}(k_{sx}=k_{0x})
\\
T_\perp  &=& S_{11}^{(1d)}(k_{sx}=k_{0x})-S_{12}^{(1d)}(k_{sx}=k_{0x}) ~~~.
\eeqa
The DDA results are in excellent agreement with the exact results
when the validity condition $|m|k_0d<0.5$ is satisfied, but results
with moderate accuracy are obtained even when $|m|k_0d\approx 1$.

\begin{figure}[t]
  \begin{center}
  \includegraphics[angle=270,width=5.0in]{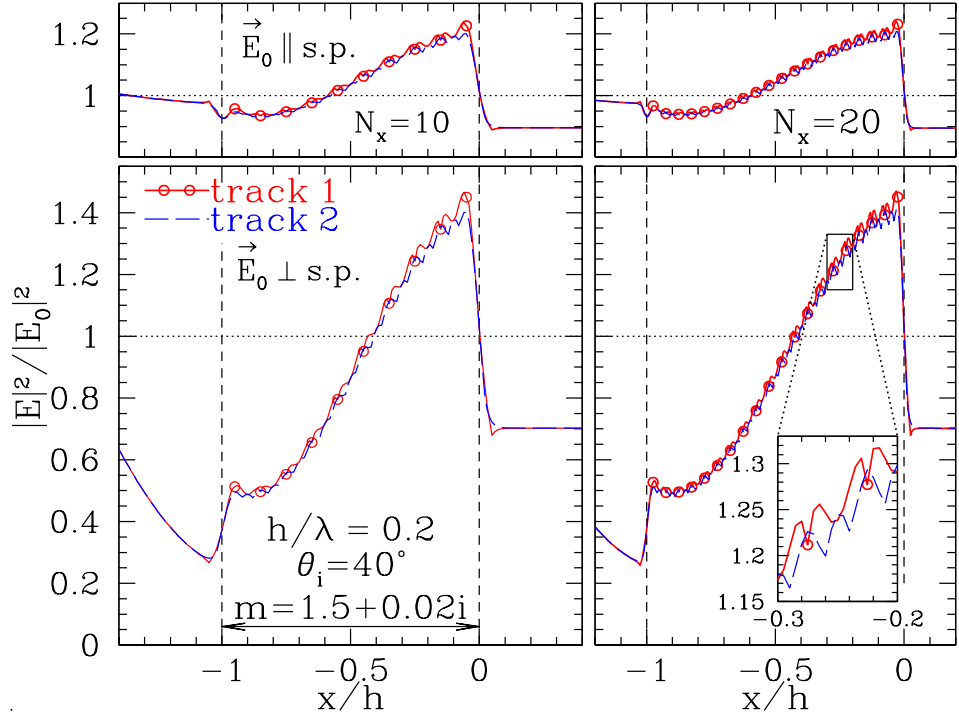}
  \caption{\label{fig:E in dielectric slab} 
     $|\bE^2|/|\bE_0^2|$ along two tracks normal to the dielectric slab of 
     Fig.\ \ref{fig:dielectric slab}b,
     for slab thickness $h=0.2\lambda$,
     incidence angle $\alpha_i=40^\circ$,
     and incident polarizations $\parallel$ and $\perp$ to the scattering
     plane (see text).
     Results were calculated using eq.\ (\ref{eq:near E})
     with the slab represented by $N_x=10$ and $N_x=20$ dipole layers
     (i.e., dipole spacing $d=0.1h$ and $0.05h$).
     The circles along track 1 are at points where dipoles are located.
     }
   \end{center}
\end{figure}

\section{Near-Field Evaluation}
The polarizations $\bP_{j00}$ can be used to calculate the
electric and magnetic fields at any point, including within or near
the target, using the exact expression for $\bE$ and $\bB$ from a point
dipole, modified by a function $\phi$:
\beqa \nonumber
\bE(\br,t)
&=& e^{-i\omega t}\sum_j{\sum_{m,n}}^\prime ~
      \frac{\exp(ik_0R_{jmn})}{|R_{jmn}|^3}
      \phi(R_{jmn})
      \bigg\{ k_0^2\bR_{jmn} \times \left(\bP_{jmn}\times\bR_{jmn}\right)
\\ \label{eq:near E}
&&    + \frac{(1-ik_0R_{jmn})}{R_{jmn}^2}
        \left[3\bR_{jmn}(\bR_{jmn}\cdot\bP_{jmn}) - R_{jmn}^2\bP_{jmn}\right]
        \bigg\}
+ \bE_0\exp(i\bk_0\cdot\br - i\omega t)
\\ \nonumber
\bB(\br,t) 
&=& e^{-i\omega t} \sum_j{\sum_{m,n}}^\prime ~ 
k^2 \frac{\exp(ik_0R_{jmn})}{R_{jmn}^2}
\phi(R_{jmn})
\left(\bR_{jmn}\times\bP_{jmn}\right)
\left(1-\frac{1}{ik_0R_{jmn}}\right)
\\ \label{eq:near B}
&&+ ~\khat_0\times \bE_0 \exp(i\bk_0\cdot\br-i\omega t)
\\
\bR_{jmn}&\equiv& \br - \br_{jmn}
\\ \label{eq:phi}
\phi(R) &\equiv&\exp\left[-\gamma(k_0R)^4)\right]\times
\left\{\begin{array}{l l}
              1 & {\rm for}~ R \geq d
\\
              (R/d)^4 & {\rm for}~ R < d
\end{array}
\right.
\eeqa
The function $\phi(R)$ smoothly suppresses the (oscillating) 
contribution from distant dipoles
in order to allow the summations to be truncated, just as
in eq.\ (\ref{eq:APBC,2d,sum}) for evaluation of $\bAtilde_{j,k}$.
If $\br$ is within the target or near the target surface, 
the summations over $(m,n)$ are limited to $|R_{jmn}| \leq 2/\gamma k_0$.
The $(R/d)^4$ factor suppresses
the $R^{-3}$ divergence of $\bE$ as $\br$ approaches
the locations of individual dipoles, and at the dipole locations results in
$\bE$ that is exactly equal
to the field that is polarizing the dipoles in the DDA formulation.
Evaluation of eq.\ (\ref{eq:near E}, \ref{eq:near B}) 
is computationally-intensive, because the summations
$\Sigma_j\Sigma_{m,n}^\prime$ typically have many terms.

To illustrate the accuracy, we consider the infinite slab of 
Fig.\ \ref{fig:dielectric slab}b, with
refractive index $m=1.5+0.02i$ and radiation
incident at an angle $\theta_i=40^\circ$.
Figure \ref{fig:E in dielectric slab} 
shows the time-averaged
$|\bE|^2/|\bE_0|^2$ for slab thickness $h=0.2\lambda$
-- near a minimum in transmission, and a maximum
in reflection (see Fig.\ \ref{fig:dielectric slab}b.
The program DDfield 
(see Appendix \ref{app:DDSCAT})
was used to evaluate $\bE$ along two lines normal to the slab: 
track~1 passes directly through dipole sites, and track~2 passes midway between
the four nearest dipoles as it crosses each dipole layer.
The E fields calculated along tracks 1 and 2 are very similar, although
of course not identical.
Within the slab, $|\bE|$ along track 2 tends to be slightly smaller than along
track 1, but for this example the difference is typically less than $\sim$1\%.
Figure \ref{fig:E in dielectric slab} shows results for the slab
represented by $h/d=N_x=10$ and 20 dipole layers
(with $|m|kd=0.19$ and 0.094, respectively).

Even for $N_x=10$, 
the electric field at points more than a distance
$d$ from the edge is 
obtained to within $\sim2\%$ accuracy at worst, which is perhaps
not surprising because, as seen in Figure \ref{fig:dielectric slab},
the calculated transmission and reflection coefficients are very accurate.
The discontinuity in $|E|^2$ at the boundary is spread out over
a distance $\sim d$.
The DDA obviously
cannot reproduce field structure near the target surface
on scales smaller than the
dipole separation $d$, but fields on scales larger than $d$ appear
to be quite accurate.  DDSCAT and DDfield should be useful tools
for studying electromagnetic fields around arrays of nanostructures,
such as gold nanodisks 
\cite{Utegulov+Shaw+Draine+etal_2007,Johnson+Kim+Utegulov+Draine_2008}.

\section{Summary}
The principal results of this study are as follows:
\begin{enumerate}
\item The DDA is generalized to treat
targets that are periodic in one or two spatial dimensions.  Scattering
and absorption of monochromatic plane waves can be calculated using algorithms
that parallel those used for finite targets.
\item A general formalism
is presented for description of far-field scattering by targets that are 
periodic 
in one or two dimensions
using scattering amplitude matrices and Mueller matrices
that are similar in form to those for finite targets.
\item The accuracy of the DDA for periodic targets is tested for two
examples: infinite cylinders and infinite slabs.  The DDA, as implemented
in DDSCAT 7, is accurate
provided the validity criterion $|m|kd \ltsim 0.5$ is satisfied.
\item We show how the DDA solution can be used to evaluate $\bE$ and
$\bB$ within and near the target, with calculations for an infinite
slab used to illustrate the accuracy of near-field calculations.
\end{enumerate}
\section*{Acknowledgments}
This research was supported in part by NSF grant AST-0406883, and by the
Office of Naval Research.  We thank Dan Mackowski for providing his code
for light scattering by infinite cylinders, 
H.~A.~Yousif for discussions concerning
scattering by infinite cylinders, 
and the anonymous referees for helpful comments.

\appendix
\section{\label{app:DDSCAT}
         DDSCAT 7 and DDfield}

The theoretical developments reported here have been implemented in a new
version of the open-source code DDSCAT
({\tt http://www.astro.princeton.edu/$\sim$draine/DDSCAT.html}).
DDSCAT 7 is written in Fortran 90, with dynamic memory allocation
and the option to use either single- or double-precision arithmetic.
DDSCAT 7 includes options for various target geometries, including
a number of periodic structures.
A program DDfield for near-field calculations is also provided.

DDSCAT 7 offers the option of using an implementation of BiCGstab
with enhancement to maintain convergence in finite precision
arithmetic
\cite{Botchev_2001}.
The matrix-vector multiplications $\bAtilde \bP$
are accomplished efficiently
using FFTs 
\cite{Goodman+Draine+Flatau_1990}.
Documentation for DDSCAT is available from ArXiv 
\cite{Draine+Flatau_2008b},
with additional information available from 
{\tt http://ddscat.wikidot.com}.

In addition to differential scattering cross sections,
DDSCAT reports dimensionless ``efficiency factors'' 
$Q_x \equiv C_x({\rm TUC})/\pi\aeff^2$ for scattering and absorption, where
$C_x({\rm TUC})$ is the total cross section 
for scattering or absorption per TUC,
normalized by $\pi \aeff^2$, where $\aeff\equiv (3V_{\rm TUC}/4\pi)^{1/3}$ 
is the radius of a
sphere with volume equal to the solid volume $V_{\rm TUC}$
in one TUC.

In the case of one-dimensional targets, with periodicity $L_y$ in
the $y$ direction, the absorption, scattering, and extinction cross
sections per unit target length are
\beq
\frac{dC_x}{dL} = \frac{1}{L_y}Q_x \pi \aeff^2
\eeq
for $x={\rm ext}$, ${\rm sca}$, and ${\rm abs}$,
where $Q_x$ are the efficiency factors calculated by DDSCAT.

In the case of two-dimensional targets, with periodicities $L_u$ and
$L_v$, the absorption, scattering, and extinction cross sections
per unit target area are
\beq
\frac{dC_x}{dA} = 
 \frac{Q_x \pi \aeff^2}{L_u L_v\sin\theta_{uv}}
~~~.
\eeq

\section{\label{app:S_i vs T_i}
         Relation Between $S_i$ and $T_i$ for Infinite Cylinders}

The analytic solution for infinite cylinders decomposes the
incident and scattered radiation into components polarized
parallel and perpendicular to planes containing the cylinder
axis and the propagation vector $\khat_0$ or $\khat_s$.
These polarization basis states differ from the choice that is
usual for scattering by finite particles, where it is customary to
decompose the incident and scattered waves into components polarized
parallel and perpendicular to the {\it scattering plane} -- the plane
containing $\khat_0$ and $\khat_s$.

In the notation of 
Bohren and Huffman \cite{Bohren+Huffman_1983}, 
the radiation scattered
by an infinite cylinder can be written
\beqa 
\left(
\begin{array}{c}
   {\bf E}_s\cdot\ehat_{s\parallel}^{(ck)}\\
   {\bf E}_s\cdot\ehat_{s\perp}^{(ck)}
\end{array}
\right)
&=&
i \exp\left(i\bk_s\cdot \br-i\omega t\right)
\left(\frac{2i}{\pi k_0 R \sin\alpha}\right)^{1/2}
\left(
\begin{array}{c c}
   T_1 & -T_3 \\
   T_3 & T_2
\end{array}
\right)
\left(
\begin{array}{c}
	{\bf E}_0\cdot\ehat_{i\parallel}^{(ck)}\\
	{\bf E}_0\cdot\ehat_{i\perp}^{(ck)}
\end{array}
\right)~~~~
\label{eq:T matrix def}
\eeqa
where $R$ is the distance from the cylinder axis,
$\alpha$ is the angle between $\bk_0$ and the cylinder axis $c$,
and superscript $(ck)$ denotes polarization vectors 
parallel or perpendicular to planes containing the cylinder axis
$\hat{\bf c}$ and either $\bk_0$ or $\bk_s$.
The azimuthal angle $\zeta$ is measured around the cylinder
axis $\hat{\bf c}$, with $\zeta=0$ for forward scattering.
The scattering angle $\theta = \arccos[\hat{\bf k}_0\cdot\hat{\bf k}_s]$ is
\beq
\theta =
\arccos\left[1-(1-\cos\zeta)\sin^2\alpha\right]
~~~.
\eeq
The scattering amplitude matrix elements $T_i$ appearing in
(\ref{eq:T matrix def}) can be related to the matrix elements $S_i$
appearing in eq.\ (\ref{eq:S for 1d periodicity}):
\beq
\bS = \left(\frac{2i}{\pi\sin\alpha}\right)^{1/2} \bA \bT \bB^{-1}
~,~~
\eeq
\beqa
\bS &\equiv&
\left(
\begin{array}{cc}
   S_2^{(1d)} & S_3^{(1d)} \\
   S_4^{(1d)} & S_1^{(1d)}
\end{array}
\right)
\hspace*{2em},\hspace*{2em}
\bT \equiv
\left(
\begin{array}{cc}
   T_1 & -T_3 \\
   T_3 & T_2
\end{array}
\right)~,~~
\\ 
\bA &\equiv&
\!\left(
\begin{array}{cc}
\ehat_{s\parallel}\cdot\ehat_{s\parallel}^{(ck)} &
\ehat_{s\parallel}\cdot\ehat_{s\perp}^{(ck)} \\
\ehat_{s\perp}\cdot\ehat_{s\parallel}^{(ck)} &
\ehat_{s\perp}\cdot\ehat_{s\perp}^{(ck)}
\end{array}
\right)
=
\frac{1}{\sin\theta}
\left(
\begin{array}{cc}
-\cot\alpha (1\!-\!\cos\theta) &  \sin\alpha\sin\zeta \\
-\sin\alpha\sin\zeta       & -\cot\alpha (1\!-\!\cos\theta) 
\end{array}
\right)~,~~
\\
\bB &\equiv&
\left(
\begin{array}{cc}
\ehat_{i\parallel}\cdot\ehat_{i\parallel}^{(ck)} &
\ehat_{i\parallel}\cdot\ehat_{i\perp}^{(ck)} \\
\ehat_{i\perp}\cdot\ehat_{i\parallel}^{(ck)} &
\ehat_{i\perp}\cdot\ehat_{i\perp}^{(ck)}
\end{array}
\right)
=
\frac{1}{\sin\theta}
\left(
\begin{array}{cc}
 \cot\alpha (1\!-\!\cos\theta) &  \sin\alpha\sin\zeta \\
-\sin\alpha\sin\zeta       &  \cot\alpha (1\!-\!\cos\theta) 
\end{array}
\right)~,~~
\\
\bB^{-1}\!\!&=&
\frac{\sin\theta}{\cot^2\alpha(1-\cos\theta)^2+\sin^2\alpha\sin^2\zeta}
\left(
\begin{array}{cc}
 \cot\alpha (1-\cos\theta) & -\sin\alpha\sin\zeta \\
 \sin\alpha\sin\zeta       & \cot\alpha (1-\cos\theta)
\end{array}
\right)~.~~
\eeqa

\begin{thebibliography}{10}
\newcommand{\enquote}[1]{``#1''}

\bibitem{Purcell+Pennypacker_1973}
E.~M. {Purcell} and C.~R. {Pennypacker}, \enquote{{Scattering and Absorption of
  Light by Nonspherical Dielectric Grains},} \apj \textbf{186}, 705--714
  (1973).

\bibitem{Draine_1988}
B.~T. {Draine}, \enquote{{The discrete-dipole approximation and its application
  to interstellar graphite grains},} \apj \textbf{333}, 848--872 (1988).

\bibitem{Draine+Flatau_1994}
B.~T. {Draine} and P.~{Flatau}, \enquote{{Discrete-dipole approximation for
  scattering calculations},} \josaa \textbf{11}, 1491--1499 (1994).

\bibitem{Draine_2000a}
B.~T. {Draine}, \enquote{{The Discrete Dipole Approximation for Light
  Scattering by Irregular Targets},} in \enquote{Light Scattering by
  Nonspherical Particles: Theory, Measurements, and Applications,} , M.~I.
  Mishchenko, J.~W. Hovenier, and L.~D. Travis, eds. (San Diego: Academic
  Press, 2000), pp. 131--145.

\bibitem{Schmehl+Nebeker+Hirleman_1997}
R.~Schmehl, B.~M. Nebeker, and E.~D. Hirleman, \enquote{Discrete-dipole
  approximation for scattering by features on surfaces by means of a
  two-dimensional fast fourier transform technique,} J. Opt. Soc. Am. A
  \textbf{14}, 3026--3036 (1997).

\bibitem{Paulus+Martin_2001}
M.~Paulus and O.~J.~F. Martin, \enquote{Green\char39{}s tensor technique for
  scattering in two-dimensional stratified media,} Phys. Rev. E \textbf{63},
  066615 (2001).

\bibitem{Yang+Liou_2000}
P.~{Yang} and K.~N. {Liou}, \enquote{{Finite Difference Time Domain Method for
  Light Scattering by Nonspherical and Inhomogeneous Particles},} in
  \enquote{Light Scattering by Nonspherical Particles: Theory, Measurements,
  and Applications,} , M.~I. Mishchenko, J.~W. Hovenier, and L.~D. Travis, eds.
  (San Diego: Academic Press., 2000), pp. 173--221.

\bibitem{Taflove+Hagness_2005}
A.~{Taflove} and S.~C. {Hagness}, \emph{{Advances in Computational
  Electrodynamics: the Finite-Difference Time-Domain Method}} (Artech House,
  Boston, 2005).

\bibitem{Markel_1993}
V.~A. {Markel}, \enquote{{Coupled-dipole Approach to Scattering of Light from a
  One-dimensional Periodic Dipole Structure},} Journal of Modern Optics
  \textbf{40}, 2281--2291 (1993).

\bibitem{Chaumet+Rahmani+Bryant_2003}
P.~C. {Chaumet}, A.~{Rahmani}, and G.~W. {Bryant}, \enquote{Generalization of
  the coupled dipole method to periodic structures,} Phys. Rev. B \textbf{67},
  165404 (2003).

\bibitem{Chaumet+Sentenac_2005}
P.~C. {Chaumet} and A.~{Sentenac}, \enquote{{Numerical simulations of the
  electromagnetic field scattered by defects in a double-periodic structure},}
  \prb \textbf{72}, 205437--20544 (2005).

\bibitem{Draine+Flatau_2004}
B.~T. {Draine} and P.~{Flatau}, \enquote{{User Guide for the Discrete Dipole
  Approximation Code DDSCAT.6.1},} {\tt http://arXiv.or/abs/astro-ph/ArXiv/0409262} (2004).

\bibitem{Draine+Goodman_1993}
B.~T. {Draine} and J.~{Goodman}, \enquote{{Beyond Clausius-Mossotti - Wave
  propagation on a polarizable point lattice and the discrete dipole
  approximation},} \apj \textbf{405}, 685--697 (1993).

\bibitem{Gutkowicz-Krusin+Draine_2004}
D.~{Gutkowicz-Krusin} and B.~T. {Draine}, \enquote{{Propagation of
  Electromagnetic Waves on a Rectangular Lattice of Polarizable Points},} 
  {\tt http://arXiv.org/abs/astro-ph/0403082}
  (2004).

\bibitem{Born+Wolf_1999}
M.~Born and E.~Wolf, \emph{Principles of Optics} (Cambridge Univ. Press,
  Cambridge, 1999).

\bibitem{Bohren+Huffman_1983}
C.~F. Bohren and D.~R. Huffman, \emph{Absorption and Scattering of Light by
  Small Particles} (Wiley, New York, 1983).

\bibitem{Utegulov+Shaw+Draine+etal_2007}
Z.~N. {Utegulov}, J.~M. {Shaw}, B.~T. {Draine}, S.~A. {Kim}, and W.~L.
  {Johnson}, \enquote{{Surface-plasmon enhancement of Brillouin light
  scattering from gold-nanodisk arrays on glass},} in \enquote{Plasmonics:
  Metallic Nanostructures and Their Optical Properties V.}, Edited by Stockman,
  Mark I.. Proceedings of the SPIE, vol. 6641 (2007), 66411M.

\bibitem{Johnson+Kim+Utegulov+Draine_2008}
W.~L. {Johnson}, S.~A. {Kim}, Z.~N. {Utegulov}, and B.~T. {Draine},
  \enquote{{Surface-plasmon fields in two-dimensional arrays of gold
  nanodisks},} submitted for publication in SPIE 2008 Optics and Photonics
  (2008).

\bibitem{Botchev_2001}
M.~A. {Botchev}, subroutine {\tt zbcg2}, 
  {\tt http://www.math.uu.nl/people/vorst/zbcg2.f90}  (2001).

\bibitem{Goodman+Draine+Flatau_1990}
J.~J. {Goodman}, B.~T. {Draine}, and P.~J. {Flatau}, \enquote{{Application of
  fast-Fourier transform techniques to the discrete dipole approximation},}
  \optlett \textbf{16}, 1198--1200 (1990).

\bibitem{Draine+Flatau_2008b}
B.~T. {Draine} and P.~{Flatau}, \enquote{{User Guide for the Discrete Dipole
  Approximation Code DDSCAT 7.0},} 
{\tt http://arXiv.org/abs/astro-ph/0809.0337} (2008).

\end{thebibliography}

\end{document}